\documentclass[floatfix,aps,prd,showpacs,amsmath,nofootinbib,preprintnumbers]
{revtex4}
\usepackage{graphicx}
\usepackage{colordvi}

\def\be{\begin{equation}}
\def\ee{\end{equation}}
\def\ba{\begin{eqnarray}}
\def\ea{\end{eqnarray}}
\begin{document}
\title{``Graceful'' Old Inflation}
\author{Fabrizio Di Marco}\email{dimarco@bo.infn.it}

\affiliation{Dipartimento di Fisica, Universit\`a  degli Studi di Bologna and INFN via Irnerio, 46-40126, Bologna, Italy}
\author{Alessio Notari}\email{notari@hep.physics.mcgill.ca}

\affiliation{Physics Department, McGill University, 3600 University Road, Montr\'eal, QC, H3A 2T8, Canada}
\date{\today}
\begin{abstract}
We show that Inflation in a False Vacuum becomes viable in the presence
of a spectator scalar field non
minimally coupled to gravity.  The field is unstable in this
background, it grows exponentially and slows down the pure de Sitter
phase itself, allowing then fast tunneling to a true vacuum.
We compute the constraint from graceful exit through bubble nucleation
and the spectrum of cosmological perturbations.
\end{abstract}

\maketitle



\section{Introduction}\label{Introduction}



Observations in the framework of the standard cosmological model tell us that
the observed Universe  is in a zero-energy vacuum state (except possibly for a
small cosmological constant which becomes relevant only at late times). From the point of view of fundamental physics, however, one can wonder why the vacuum energy is zero.
The question that we want to answer in this paper is the following: is it
possible that the Universe started in a different vacuum state with large
zero-point energy and then tunneled to the zero-energy state (which could very reasonably the minimum available energy state)?
\footnote{It is even more appealing if we follow Coleman and De Luccia arguments \cite{Coleman}, who concluded that in the thin wall limit and for large enough bubbles, tunneling is suppressed to a negative energy state, or Banks \cite{Banks} who claims that such a tunneling cannot occur since instantons that interpolate to a negative energy vacuum simply do not exist in a consistent theory.}

At the same time it is also
well-known that a Universe with a large vacuum energy inflates. And, in fact,
this was the first way to introduce Inflation (``Old Inflation", Guth,
1982~\cite{Guth}). As Guth himself already realized in the original paper, the
simplest version of this idea does not work, since inflation does not end
successfully (``graceful exit" problem).

A mechanism that could  realize this
idea (often called False Vacuum Inflation or also First-Order Inflation, since it would end through a first
order transition with nucleation of bubbles of true vacuum) in a simple way,
appears very attractive: in fact, it would explain dynamically why the
Universe has almost zero cosmological constant, and at the same time it would
provide Inflation, without the need for a slow-rolling field. In this paper we
show that all this is possible with the addition of {\it one simple
  ingredient}: the presence of a non minimally coupled scalar field.

This
model has already been proposed by Dolgov \cite{Dolgov} in 1983 in order to
provide a solution to the cosmological constant problem: the scalar field is
able to screen the bare value of $\Lambda$ at late times. However the model
was not a realistic description of our Universe, since at late times the
Newton constant was driven to unacceptably small values. Our idea instead is
to use this model in order to explain how a False Vacuum Inflation can end
successfully through Bubble Nucleation.
 A model in the very same spirit was proposed already in the '80s,
 known as Extended Inflation (EI) \cite{johri, extended}.
The problem with Old Inflation was that, in order to have sufficient inflation
one needs  the ratio $r=\Gamma_{\rm{vac}}/H_I^4$ (where
$\Gamma_{\rm{vac}}$ is the decay rate per unit volume of the true vacuum to
the false vacuum and $H_I$ is the Hubble constant during inflation) to be much
smaller than 1. At the same time Inflation can end only if $r$ becomes of
order one. So only a model with variable (increasing) $r$ can be viable. The
idea of Extended Inflation was to consider a Brans-Dicke gravitational theory,
in which the vacuum energy does not produce de Sitter Inflation but Power-Law
Inflation. This gives indeed an increasing $r$ and so successful completion of
the phase transition. However, the model had a definite prediction: the
spectral index of cosmological perturbations had to be roughly $n_S\lesssim
0.8$, and this was ruled out in 1992 by the Cosmic Background Explorer
satellite (COBE) experiment on the
anisotropies the Cosmic Microwave Background \cite{Liddle}, so many
variants of this model were proposed \cite{hyperextended,
  Barrow-Maeda}.
Another possibility for making $r$ variable that has been explored is having $\Gamma_{\rm{vac}}$ variable through the introduction of a slowly rolling field \cite{double}.
Our proposal still keeps $\Gamma_{\rm{vac}}$ constant, but it differs from the so-called hyperxetended \cite{hyperextended,
  Barrow-Maeda} models, since we introduce an initial de Sitter stage and then we show that this stage dynamically evolves in a second stage, which is the same as in the original Extended Inflation. In this sense our proposal is as simple as the original EI: the difference is that it has two different periods of inflation.

The first period is exactly de Sitter and it is
the period in which the spectrum of perturbations is produced on the scales
relevant to the observations (surprisingly none of the modifications of
Extended Inflation has considered the simple possibility of having an early de Sitter stage, the only mention being in \cite{LiddleWands91}, which however did not consider it as a viable opportunity). During this period the non minimally coupled
scalar field grows exponentially from small to large values. The second period
instead starts when the non minimally coupled scalar becomes important, it slows down inflation dramatically
and it leads to a power-law behaviour (which could even be decelerating). So it can lead to a successful transition to
the radiation era through bubble nucleation.

The plan of the paper is as follows. In the next section we introduce the model proposed by Dolgov~\cite{Dolgov},
summarizing the dynamics of the classical homogeneous fields. In
section~\ref{bubble} we analyze the constraints on the parameters of the model
in order to give sufficient inflation and graceful exit, both analytically and
numerically. In section~\ref{spectrumSect} we study the production (spectral index
and amplitude) of density fluctuations in this minimal scenario, and we also
discuss the fluctuations due to an additional minimally coupled scalar field
which may act as a curvaton. In section~\ref{einstein} instead we show possible ways to
recover Einstein gravity at late times. Section~\ref{other} contains a
discussion of other possible effects and of other observables still to be
computed, that are not covered in this paper. Finally, in
section~\ref{conclusions} we draw our conclusions. The paper contains also two
appendices. Appendix~\ref{BubblesApp} contains the calculation of the
constraint coming from the absence of detection of big bubbles in the
CMB. Appendix \ref{Amplitude} contains some calculations of the initial amplitude of the field $\phi$
and of the perturbation in the energy density due to fluctuations in the non minimally coupled field.



\section{The model}\label{model}



As we have mentioned in the Introduction, the model that we are going to study has an unstable vacuum energy, that we will call $\Lambda$, with some tunneling rate per unit volume $\Gamma_{\rm{vac}}$\footnote{Note that, although we are using the name $\Lambda$, this is not a cosmological constant, but a vacuum energy that can decay with some rate $\Gamma_{\rm{vac}}$, as happens for example for a scalar field trapped in false vacuum. This can be in principle computed by knowing the details of the potential barrier that separates the two vacua.}.

The only other ingredient of our model at this point is, as in \cite{Dolgov}, a scalar field $\phi$ which has a generic (``non-minimal'') coupling to the Ricci scalar $R$.
So, the action is:
\be
 S = \int d^4 x \sqrt{-g} \left[  \frac{1}{2} (M^2 + \beta \phi^2 )R - \frac{1}{2} \partial_{\mu}\phi \partial^{\mu}\phi
 -\Lambda \right] \label{lagrangiana} \, ,
\ee
where it is crucial to assume $\beta>0$.
The reader should note that the field $\phi$ is a spectator and it has nothing to do with the false vacuum energy $\Lambda$.

Note also that the bare mass in the Lagrangian $M$ does not coincide in general with the value of the Planck mass that we observe today ($M_{\rm{Pl}}$).
The Einstein equations that follow from this action\footnote{In this paper we
  stick always to the Jordan frame. For the analysis in the Einstein frame see
  the  case $n=2$ of \cite{tirtho}.} in the homogeneous case are:
\begin{eqnarray}
H^2 &=& \frac{1}{3 (M^2 + \beta \phi^2) }\left[ \frac{1}{2} \dot\phi^2 - 6 H \beta \phi \dot\phi + \Lambda \right] \, , \\
\dot H &=& - \frac{1}{2(M^2 + \beta \phi^2 )}\left[ \dot\phi^2 - 8 H \beta \phi \dot\phi + 2 \beta \dot\phi^2 + 2  \beta^2 \phi^2 R \right] \, ,
\end{eqnarray}
where $H\equiv\dot{a}/a$ ($a$ is the scale factor), and the Ricci scalar is:
\be
R=6 \dot{H}+12 H^2 \, ,
\ee
and the evolution equation for the scalar field is:
\be
 \ddot\phi + 3H \dot\phi - \beta R \phi = 0 \, .
\ee


The energy density and pressure for the field $\phi$ are:
\begin{eqnarray}
\rho_{\phi} &=&   \frac{1}{2} \dot\phi^2 - 6 H\beta \phi \dot\phi - 3\beta H^2  \phi^2, \\
 p_{\phi} &=& \frac{1}{2} \dot\phi^2 - 2 H \beta \phi \dot\phi + 2 \beta^2 R
 \phi^2 + 3 H^2 \beta \phi^2 + 2 \dot H \beta \phi^2 + 2\beta \dot\phi^2 \, .
\end{eqnarray}

Now, let us assume that the Universe starts in the false vacuum, and that the field $\phi$ sits close to zero at the beginning.
So, at initial time the evolution of the scale factor is given by a de Sitter phase:
\be
H^2 \simeq H_I^2 \equiv \frac{\Lambda}{3 M^2}  \, .
\ee
The equation of motion for $\phi$ in such a background is:
\be
\ddot\phi + 3H_I \dot\phi - 12 H_I^2 \beta\phi = 0  \, ,
\ee
and its growing solution is:
\be
\phi(t) = \phi_0 e^{(\epsilon H_I t)/2 } \, , \hspace{1cm} \epsilon \equiv
3\left(-1 + \sqrt{1 + \frac{16}{3} \beta} \, \right) \qquad
\left(\epsilon \simeq 8 \beta \, \, \,
{\rm for \,\, small\,\,} \beta \, \right)  \label{growing} \, .
\ee
As we said, we are going to choose the initial condition $\phi_0$ close to
zero (and zero initial $\dot{\phi}_0$). Actually it is not physical to assume
that $\phi$ starts exactly in zero, since the field is subject to quantum
fluctuations, and so the initial value is of the order $\phi_0\approx H_I$
(see Eq.~(\ref{fi0}) for a more precise estimate).
So the field $\phi$ is unstable in this background, and indeed it drives the de Sitter phase itself to an end. In fact, the energy density related to the field $\phi$ is negative:
\be  \rho_{\phi} \simeq - A\,H_I^2\, \phi_0^2 \, e^{\epsilon H_I t} \, ,\hspace{1cm}  A
\equiv  3 \beta(1 + \epsilon) - \frac{1}{8} \epsilon^2 \,>\,0 \, ,\ee
and increases with the time in magnitude.

After a period of transition, at late times the field $\phi(t)$ becomes linear
in $t$, its energy density becomes constant, and the scale-factor increases with time as a power-law:
\be a(t) \sim t^{\alpha} \, , \hspace{2.5cm} \phi(t) \sim B t \, , \ee
where:
\be \alpha \equiv \frac{1+ 2 \beta}{4 \beta} \, , \hspace{1.cm}
B \equiv \frac{4\sqrt{\beta\Lambda}}{\sqrt{60 \beta^2 + 28 \beta + 3}}\, .\ee

\begin{figure}
\includegraphics[width=0.45\textwidth]{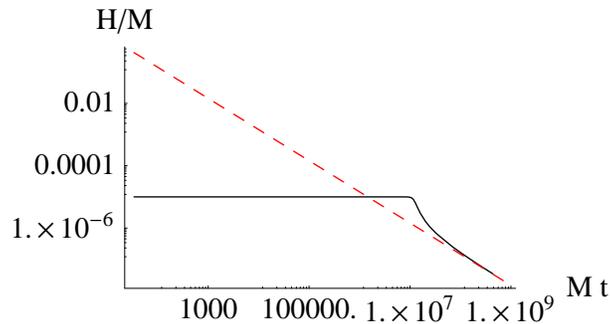}\caption{\label{figuraHdit}
The evolution of the Hubble parameter as a function of time in our model (solid black line), compared to a power-law evolution $H=\alpha/t$ (red dashed line). Here $\beta=1/56, H_I=10^{-5}M$.
}
\end{figure}

A numerical solution of the system is presented in FIG. \ref{figuraHdit}
\footnote{ Some readers might feel uncomfortable with $\phi$ becoming bigger than the fundamental scale $M$, being worried about quantum gravitational corrections. However these are well known to be under control \cite{LindeBook} if the energy density and the masses are below $M$, which is the case here.}.



This late time behaviour is due to the fact that, when $\phi$ becomes very large, the initial ``bare'' mass $M$ becomes subdominant, and so the system reduces to a pure Brans-Dicke model of gravity. It is known that vacuum energy in Brans-Dicke theory leads to power-law expansion: this was in fact used as a proposal for the graceful exit in the Extended Inflation scenario.
The crucial difference here is that we have a dynamical transition between Einstein gravity (that leads to the exponential phase of inflation) and the Brans-Dicke gravity (that leads to the power law phase of inflation).
We will exploit the presence of the first phase in order to produce a flat spectrum of cosmological perturbations, and the presence of the second phase in order to have a graceful exit from inflation to the radiation era: in fact $H$ decreases and when it becomes of the order of the decay rate $\Gamma_{\rm{vac}}^{1/4}$ it allows the system to tunnel successfully to the true vacuum.



\section{Constraints on Inflation}\label{bubble}



We show in this section under which conditions the model provides a ``graceful exit'' from inflation.
The transition from the ``false'' vacuum to the ``true'' vacuum with zero energy density happens through bubble nucleation.
The transition has to be sufficiently abrupt, so that essentially all of the
bubbles are created in a very short time. The nucleation in the early stages
of inflation, in fact, would produce bubbles that would be stretched to very
large scales by the subsequent inflationary phase, and that would spoil the
observed isotropy of the CMB on large scales \cite{weinberg, La, LiddleWands} (``Big Bubbles'' constraint).


First, we introduce the following definition for the number of e-folds:
\be
N(t)=\int^{t_{\rm end}}_{t} H(\tilde{t}) d\tilde{t}  \label{efolds}  \, ,
\ee
where $t_{\rm end}$ is the time at which inflation ends through bubble nucleation and
the radiation era starts. Assuming that inflation starts at $t=0$, we
define $N_{\rm tot} \equiv N(0)$ as the total number of e-folds in inflationary
stage. Our model can be successful if it is able to produce the correct
spectrum of perturbations on large scales and, as it is well-known, a flat spectrum
can be achieved if $H$ is sufficiently constant during inflation. This seems
to be in contrast with the requirement that in our model $H$ has to decrease
in order to allow the phase transition. However, the crucial point that we
want to stress here is that the requirement that we need in order to satisfy
observations is not so strong: we only know that the scales between $3000
h^{-1}{\rm Mpc}$ and about $ 50 h^{-1}{\rm Mpc}$ have a flat spectrum \cite{MAP}. This can be achieved if these scales go out of the horizon during
the first stage of inflation, in which the Hubble parameter is constant. In
usual inflationary scenarios, these scales are produced roughly between
$60\lesssim N \lesssim 64$. For the following we define $N_{L} \equiv N(t_L) $, where
$t_L$ is the time at which a given scale $L$ crosses outside the horizon.
We also define the {\it phase I} as the phase in which $H=H_I$ is a constant,
going from the initial time $t=0$ up the time $t_{I}$ where a
transition happens. In terms of e-folds it goes on for  $N_I \equiv N_{\rm tot} -
N(t_I)$ . The {\it phase II} starts at $t_{II}$ (slightly after $t_I$), when $H$ begins to evolve
such $\alpha / t$, and ends when the vacuum decays, at a time $t_{\rm end}$, and
goes on for $N_{II} \equiv N(t_{II})$ e-folds. There are three basic
requirements that we must satisfy:\\
\\
\noindent
$\bullet\,$ on the relevant scales ({\it i.e.} during the {\it phase I} ) the production of bubbles of true vacuum has to be suppressed, and the only source of perturbations have to be the usual quantum fluctuations of scalar fields,\\
\noindent
$\bullet\,$ the {\it phase II} has to start after the production of perturbations on
these scales: $ N_{II}\lesssim N_{50 h^{-1}{\rm Mpc}}$,\\
\noindent
$\bullet\,$ inflation has to be sufficiently long: $N_{\rm tot}\gtrsim N_{3000
    h^{-1} {\rm Mpc}}.$\\
\noindent

The crucial quantity that regulates the production of bubbles is:
\be
r(t)\equiv \frac{\Gamma_{\rm{vac}}}{H(t)^4}\, .
\ee
The end of inflation $t_{\rm end}$ is achieved when this quantity is of order 1. More precisely \cite{Turner} the condition for the percolation of bubbles is:
\be
r(t_{\rm end})=\frac{9}{4 \pi} \, .
\ee

\begin{figure}

\includegraphics[width=0.45\textwidth]{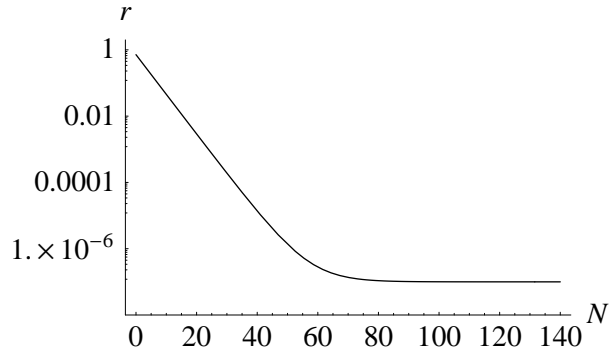}

\caption{\label{figuraHdiN} The evolution of $r\equiv \Gamma_{\rm{vac}}/H^4$ as a function of the number of e-folds N (with $\beta=1/56,\, H_I=10^{-5} M, \, r_0=10^{-7}$). Time goes from right to left: $r=r_0$ at the beginning and then it grows to ${\cal O}(1)$.
}\end{figure}

We show in FIG. \ref{figuraHdiN} the typical behaviour of $r(N)$ which is clearly constant for large $N$ and it increases with decreasing $N$ up to $\frac{9}{4 \pi}$ (where $N$ is defined to be zero).


The value of the ratio during the {\it phase I} is a free parameter, but it is constrained to be small:
\be
r_0=\frac{\Gamma_{\rm{vac}}}{H_I^4}\lesssim 10^{-7}  \label{conditionr0} \, ,
\ee
in order to prevent big bubbles formation. As we mentioned, in fact, a bubble nucleated so early would have very big size after inflation and it would appear as a black patch in the microwave sky. The number given as a bound is computed in Appendix~\ref{BubblesApp}.

So, what is important is to constrain the ratio $r(N)$ as a function of the number of e-folds.

In order to give a quantitative constraint we need the correspondence between a scale $L$ and the $N$ at which they are produced ($N_{L}$), which is given in general by:
\be
L \left(\frac{T_0}{T_{\rm{rh}}} \right)e^{- N_{L}}=H_{L}^{-1} \,
 ,
\ee
where $H_{L}$ is the value of the Hubble constant when the scale $L$ exits the
horizon, and  $T_{\rm {rh}}$ is the temperature at the beginning of the
radiation era. So, the relation is:
\be
N_{L}= 63.3 + \Delta N
+\ln\left[\frac{L}{3000 \,h^{-1}\rm Mpc} \right]  \, .
\ee
Since we assume that the scales between $3000 h^{-1}{\rm Mpc}$ and about $ 50 h^{-1}{\rm Mpc}$ are produced during the {\it phase I}, we get the following two relations:
\be
N_{3000 h^{-1}\rm{Mpc}} \simeq \left( 63.3 + \Delta N \right) \,, \qquad
N_{50 h^{-1}{\rm {Mpc}}}\simeq \left( 59 + \Delta N \right) \, , \label{dueN} \ee
where:
\be
\Delta N \equiv 3.7 + \ln\left[\frac{H_{I}}{T_{\rm rh}}  \right] \, . \ee
In our model we may assume that reheating is almost istantaneous due to rapid bubble collisions, and so:
\be
T_{\rm{rh}}^4=\Gamma_{\rm{vac}}^{1/2} (M_{\rm{Pl}}^{(\rm R)})^2= r_0^{1/2} H_I^2
(M_{\rm{Pl}}^{(\rm R)})^2 \, ,
\ee
where the superscript ``R'' means that it is the effective value of the Planck mass at the beginning of the radiation era. This implies:
\be
\Delta N\simeq \frac{1}{2} \ln\left[\frac{H_{I}}{10^{14} \rm{GeV}}\right]
-\frac{1}{8} \ln\left[\frac{r_0}{10^{-7}}\right]-\frac{1}{2} \ln\left[
  \frac{M^{(\rm R)}_{\rm {Pl}}}{10^{19} \rm{GeV}}\right]  \, .
\ee
Since the amplitude of the cosmological perturbations constrains $H_I$ to be
at most $10^{-5} M$ (see Eq.~(\ref{necessary})), the biggest value that we can get for $\Delta N$ is around:
\be
\Delta N\simeq -\frac{1}{8} \ln\left[\frac{r_0}{10^{-7}}\right]  \, . \label{biggest}
\ee

\subsection{Analytical approximation}

We give here an understanding of all the constraints showing their dependence
on the parameters of the model with an analytical approximation. A more
precise numerical analysis of the ratio $r(N)$ is provided numerically in the
next subsection. The number of e-folds of the first phase is given by:
\be
N_I  \simeq H_I t_{I} \,
\ee
where the time $t_{I}$ of the end of the first phase is found equating the energy density in the field $\phi$ to the bare vacuum energy $\Lambda$:
\be
\Lambda \,\simeq \, \left| \rho_{\phi}(t_I) \right| \, = \, A H_I^2 \phi(t_{I})^2 \, .
\ee
Using  Eq.~(\ref{growing}) we have:
\be
t_{I}\simeq\frac{1}{\epsilon H_I} \ln\left[\frac{\Lambda}{A H_I^2 \phi_0^2}\right] \, ,
\ee
and so the number of efolds is:
 \be
N_I \simeq \frac{1}{\sqrt{9+48 \beta}-3} \ln\left[\frac{4 M^2}{\phi_0^2
   \left (\sqrt{9+48 \beta}\,(1 + 4 \beta) - 3 - 16 \beta \right)     }\right]  \label{NI} \, .
\ee
It is of some utility to expand it for small $\beta$:
\be
N_I\simeq  \frac{1}{8 \beta} \left(2 \ln\left[\frac{M}{\phi_0}\right]- \ln[\beta] \right) \, .
\ee


It is in principle more difficult to estimate the duration of the remaining part of
inflation, since  we do not have an analytical expression for the transition between the {\it phase I} and {\it phase II} , when $H$ drops from the value $H_I$ to $\alpha/t$.
However, starting from the end of inflation we can use the asymptotic solution
$H=\alpha/t$ and so we get from Eq.~(\ref{efolds}):
\be
N(t)=\alpha \, \ln\left[\frac{t_{\rm end}}{t}\right] \, .
\ee
We have also that in the asymptotic phase the ratio $r(N)$ goes as:
\be
r(N)=\frac{\Gamma_{\rm vac}}{\alpha^4}\, t_{\rm end}^4 \, e^{-4\frac{N}{\alpha}}=
\frac{9}{4 \pi} \, e^{-4\frac{N}{\alpha}} \label{analitic}  \, .
\ee
We can roughly use this solution to find when the {\it phase II}  begins:
\be
\frac{9}{4 \pi} \, e^{-4 N_{II}/\alpha}\, \approx \, r_0 \, ,
\ee
that leads to:
\be
N_{II}\simeq - \frac{\alpha}{4} \ln\left[ \frac{4 \pi}{9} r_0  \right] \, . \label{NII}
\ee
So, imposing $N_{II}<N_{50 h^{-1} \rm Mpc}$ we have a constraint on $\alpha$ (at given $r_0$):
\be
\alpha \lesssim -4(59+\Delta N)/\ln\left[\frac{4\pi r_0}{9}\right] \,. \label{boundr0}
\ee
Now inserting here the maximal value for $\Delta N$ of Eq.~(\ref{biggest}), we obtain:
\be \alpha \lesssim \frac{-236 + \frac{1}{2} \ln\left[\frac{r_0}{10^{-7}}
\right] }{\ln[\frac{4\pi r_0}{9}]} \, .\ee
This translates in a constraint on $\beta$ (or equivalently on $\omega\equiv\frac{1}{4 \beta}$ which is sometimes used in the literature). Assuming the maximal possible value $r_0 = 10^{-7}$ we get:
\be
\beta \gtrsim \frac{1}{58}  \qquad  {\rm or} \qquad  \omega \lesssim 14.5 \, .  \label{boundanalitico}
\ee

As we said this number is only an estimate, and we have the correct constraint
coming from the numerical analysis, in Eq.~(\ref{boundnumerico}).
It is relevant to stress that this condition is independent on the third parameter of our model, that is the value of $H_I/M$ (or equivalently the value of the vacuum energy $\Lambda$).
As we will show in section \ref{spectrumSect}, the range of $\beta$ that we get will severely constrain the spectral index of the fluctuations in the field $\phi$.

Then, the third constraint that we have to impose is that the duration of
inflation is sufficiently long. In terms of e-folds, the duration of the first
phase is given by Eq.~(\ref{NI}), while the duration of the second phase is given by Eq.~(\ref{NII}), so one can check for a given
set of the parameters whether $N_I+N_{II}\gtrsim N_{3000 h^{-1} \rm
  {Mpc}}$\footnote{Here we are disregarding the transition between {\it phase
    I} and {\it II}, so that $N_{\rm tot}\approx N_I+N_{II}$.} is satisfied or not. There are three relevant parameters for this condition: $\phi_0/M$ (which is actually $H_I/M$ if the initial condition is given by quantum fluctuations), $\beta$ and $r_0$. The condition is very easy to satisfy for small $\beta$ and more difficult for $\beta$ approaching the value 1/2 (for $\beta>1/2$ we do not have inflation anymore in the {\it phase II} ).



\subsection{Numerical analysis} \label{numerical}



In this section we present the result from the numerical solution of the differential equations that govern our system. In particular we are interested in $r(N)$ and in constraining it.

The way in which we applied the conditions on the duration of inflation given above is as follows:
\begin{itemize}
\item we compute $r(N)=\Gamma_{\rm vac}/H(N)^4$ and impose that $r(N_{50
    h^{-1} \rm Mpc})<10^{-7}$,
\item we impose that the variation of $H$ between $N_{50 h^{-1} \rm Mpc}$ and
  $N_{3000 h^{-1} \rm Mpc}$ is small \footnote{By ``small'' we mean here an amount that does not change substantially the spectral index. Precisely we take $2 \Delta H/\Delta N (1/H)\lesssim 0.1$, where $\Delta H\equiv H(N_{3000 h^{-1}{\rm Mpc}})-H(N_{50 h^{-1}{\rm Mpc}})$. In fact this is the correction that would get a spectral index in slow-roll approximation, and we want it to be smaller than $0.1$.},
\item we check that $N_{\rm tot}\gtrsim N_{3000 h^{-1}\rm Mpc}$.
\end{itemize}

\begin{figure}

\includegraphics[width=0.45\textwidth]{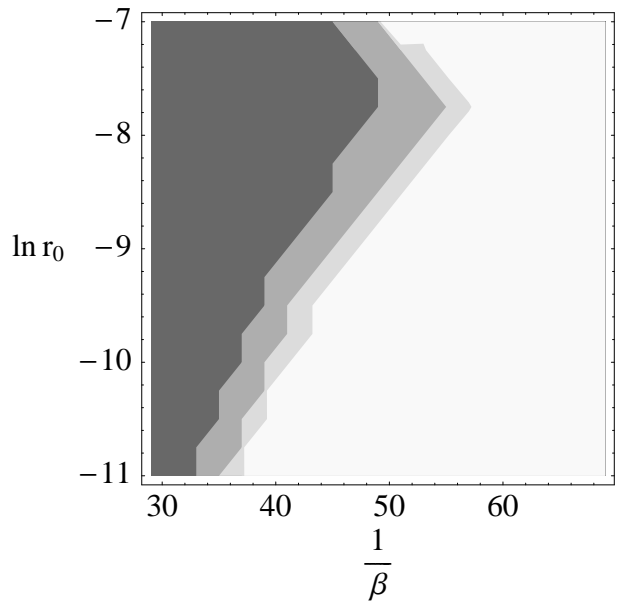}

\caption{\label{plotfinal}
In grey we show the regions of allowed parameter space, for different $\Delta N$ (from left to right $\Delta N=-8,-4,0$).
The region on the top and the region on the right are excluded because of the constraint on the production of large bubbles (this is independent on the value of $H_I/M$).
For $\beta$ becoming bigger than what is shown in the plot, the scenario is still viable except when $\beta$ gets closer to ${\cal O}(0.1)$, where sufficient inflation might become difficult to obtain (depending also on the value of $H_I/M$). }

\end{figure}

The region of parameters in which the model works is shown in FIG. \ref{plotfinal} for some different values of $\Delta N$.

As we showed in the previous subsection, there are two main parameters: the
ratio $\Gamma_{\rm {vac}}/H_I^4$  and the nonminimal coupling parameter
$\beta$. As we have shown qualitatively in Eq.~(\ref{boundr0}), the viable
interval in the parameter $\beta$ shrinks as $r_0$ decreases. This is respected by the
behaviour of FIG. \ref{plotfinal}, except for the upper part of the figure,
where the relevant scales are produced exactly at the transition between
exponential inflation and power-law, and so $r(N)$ decreases slower than our
analytical approximation $r_A(N)$ of Eq.~(\ref{analitic}).

The most important information that appears from the exclusion plot is that the mechanism works with an upper bound on $\beta$. Considering that $\Delta N\lesssim 0$, from the plot we can see that:
\be
\beta \gtrsim \frac{1}{57} \, ,   \label{boundnumerico}
\ee
which means that the analytical approximation Eq.~(\ref{boundanalitico}) already gave a good estimate.

Also, note that the dependence on $H_I$ comes only in two places. First, $H_I$
enters in defining what is $\Delta N$. Second, when checking if the model is
able to give sufficient inflation (which is not shown in FIG. \ref{plotfinal}). In fact, the duration of the exponential phase is
related to the ratio $\phi_0/M$ (see Eq.~(\ref{NI})), and $\phi_0\approx H_I$
if we assume that $\phi_0$ is set by the quantum fluctuations (see Eq.~(\ref{fi0})).
It becomes difficult to have sufficient inflation only for relatively big values of $\beta$ (close to ${\cal O}(0.1)$) and when $H_I$ is close to the scale $M$.


\section{Spectrum of perturbations}\label{spectrumSect}



In this section we compute the spectrum of cosmological perturbations produced in our scenario, in order to see if it is consistent with the observations.

The perturbations on relevant cosmological scales are produced in our model during the phase of exponential expansion.
For this reason one can realize situations in which the observed spectrum is flat, {\it i.e.} $n_S=1$.

For comparison Extended Inflation \cite{extended} instead has always a power
law $t^{\alpha}$ expansion , where the Big Bubbles constraint requires
$\alpha<20$ as we discussed in sect.~\ref{bubble} \cite{LiddleWands}. The constraint on $\alpha$ produced a spectral index $n_S\lesssim 0.75$ for any fluctuating field, which was clearly ruled out by observations, after COBE \cite{Liddle}.
Even adding a curvaton did not help \cite{liberated}, since any field has generically a non-flat spectrum, due to the fact that $H$ is never constant.

In our model, first we analyze the slope and the amplitude of the spectrum of
perturbations due to fluctuations of the field $\phi$.  Physically the fluctuations in $\phi$ (which is responsible for ending inflation) give rise to end of inflation at different times in different Hubble patches, therefore providing a mechanism for density perturbations.
However, even with the presence of our exponential phase, this minimal
scenario appears to be in disagreement with the observed spectrum. The reason is that the field $\phi$ is not minimally coupled and so its
fluctuations do not have $n_S=1$ even if the Hubble constant is exactly
constant during the first phase of inflation. As we will show the situation is
improved with respect to Extended Inflation (precisely $n_S-1$ is a factor
of 2 closer to 1), but the Big Bubbles constraint still constrains $n_S$ to be
too small, using the WMAP data.

One possibility (which is under study \cite{tirtho} and it is not developed
in this paper, except for some remarks in sect.~\ref{other}) to construct an inflationary model fully compatible with observations is to modify the original Lagrangian in order to obtain a strong
successful transition even with a small $\beta$ during the {\it phase I}.
Another possibility is to keep our Lagrangian, noting that a good spectrum can
easily be produced assuming that another light minimally coupled field (so,
with flat spectrum of fluctuations during the exponential phase) is
responsible for the generation fo the curvature perturbation. We will show the
viability of this possibility in subsection \ref{curvaton}.

\subsection{Computing spectral index and amplitude with only $\phi$} \label{onlyphi}

In order to get the spectrum of the cosmological perturbations we consider fluctuations
$\delta\phi(x,t)$ associated to the scalar field $\phi$:
\be \phi(x,t) = \phi(t) + \delta\phi(x,t) \, ,\ee
and we follow the treatment of cosmological perturbations in generalized gravity
theories given by \cite{Hwang,Hwang2}. In a flat cosmological model with only
scalar-type perturbations:
\be ds^2 = - (1 + 2 \psi ) dt^2 - \chi_{,\mu}\, dt\, dx^{\mu} + a^2 (1 + 2
\varphi)\delta_{\mu \nu}\, dx^{\mu}dx^{\nu} \ee
we may study the evolution of the fluctuations by means of the gauge-invariant combination:
\be
\delta\phi_{\varphi} = \delta\phi - \frac{\dot\phi}{H} \varphi \, , \ee
which is $\delta\phi$ in the uniform-curvature gauge ($\varphi = 0$). The
equation of motion for $ \delta\phi_{\varphi}$ is:
\be
\delta{\ddot\phi_{\varphi}} + \frac{(a^3 Z)^{.}} {a^3 Z}\delta{\dot\phi_{\varphi}} - \left\{\frac{1}{a^2} \nabla^2 + \frac{H}{a^3 Z \dot\phi}\left[a^3 Z\left(\frac{\dot\phi}{H}\right)^{.}\right]^{.}\right\}\delta\phi_{\varphi}
= 0 \, , \label{deltaphi} \ee
where the function $Z$ is defined as:
\be Z(t) = \frac{(M^2 + \beta \phi^2)(M^2 + \beta \phi^2 + 6 \beta^2 \phi^2)}{(M^2 + \beta \phi^2 + \beta \phi(\dot\phi / H))^2} \,.
\ee
In the large-scale limit the solution is:
\be  \delta\phi_{\varphi} = - \frac{\dot\phi}{H} \left[C(x) -D(x)\int{\frac{H^2 }{a^3 Z \dot\phi^2}}dt\right]\, , \label{CD}
\ee
and, ignoring the decaying mode $D(x)$, we
can relate the spectrum of $\delta\phi_{\varphi}$ to the
time independent function $C(x)$ which represents the perturbed 3-space
curvature in the uniform curvature gauge. So, we can compute the amplitude of
scalar perturbations by means of the amplitude of the variable $C(x)$. \par
The next step is to rewrite Eq.~(\ref{deltaphi}) (in $k$-space and in
conformal time $\tau$) in terms of the variable $ v \equiv a \sqrt{Z}
\delta\phi_{\varphi} $, and we can take into account
that, during the {\it phase I} (in which the perturbations are produced), the function $Z$
reduces to  $Z \simeq 1 $.
So
the equation of motion becomes:
\be
v''+\left( k^2 - \frac{2}{\tau^2} (1+6\beta)  \right) v=0  \label{eqv}\, . \,
\ee
 This result agrees with the general fact that a conformally coupled field ($\beta=-\frac{1}{6}$) has only Minkowski space fluctuations.
We can define the conventional constant $\nu$ as:
\be \nu \equiv\frac{3}{2}+\frac{\epsilon}{2}  \, ,
 \label{nu} \ee
which, for small $\beta$, becomes:
\be
\nu\simeq\frac{3}{2} + 4 \beta.
\ee
The solutions of Eq.~(\ref{eqv}) are:
\be v(k, \tau) =  \frac{1}{2}\sqrt{\pi\left|\tau\right|} \left[
  c_1(k)H_{\nu}^{(1)}(k,|\tau|) +  c_2(k) H_{\nu}^{(2)}(k,|\tau|)\right] \, , \ee
where $c_1(k)$ and $c_2(k)$ are constrained by the quantization conditions  $ |c_2(k)|^2 - |c_1(k)|^2 = 1$ and $H^{1,2}_{\nu}$ are the Hankel functions of the first and second kind.
So, on large scales the power-spectrum of $C$ is:
\be
P^{1/2}_C(k,\tau) = \left|\frac{H}{\dot\phi}\right| P^{1/2}_{\delta\phi_{\varphi}}(k,\tau) = \left|\frac{H}{\dot\phi}\right| \frac{H}{2 \pi} \frac{\Gamma(\nu)}{\Gamma\left(\frac{3}{2}\right)}
\left[\frac{1}{2} \frac{k}{aH}\right]^{\frac{3}{2}-\nu}\left|c_1(k) -
 c_2(k)\right|. \label{spectrum} \ee
>From here we can immediately extract the spectral index $n_S$ for scalar
perturbations (using the Bunch-Davies vacuum $c_2 =1$):
\be
n_S \equiv 1 + \frac{d \ln P_C}{d \ln k } = 4 - 2 \nu=1-\epsilon \, ,
\ee
and for small $\beta$ this becomes:
\be
n_S \simeq 1 - 8 \beta \, .
\ee
As we said, we have $\beta\gtrsim 1/57$, so $n_S\lesssim 0.86$. Unfortunately
this is not in agreement with recent observations (such as Wilkinson Microwave
Anisotropy Probe (WMAP) \cite{MAP}); therefore the model as we presented requires some corrections in order to work, as we said, modifying the Lagrangian or adding a curvaton.

Let us compute the amplitude of the perturbations in $\delta\phi$, so that we
will impose that it is suppressed (in case we add a curvaton this is required). The
amplitude of the spectrum is:
\be
 A_{C} =\left|\frac{H_I}{\pi \, \epsilon \, \phi}\frac{\Gamma(\nu)}{\Gamma(\frac{3}{2})}\right|^2 \,\,
\label{powerzeta}\, .
\ee
This has to be evaluated at $N_{3000 h^{-1}{\rm Mpc}}$, therefore $\phi$ at that time has to be much larger than $H_I$, if we want the amplitude of the perturbations to be small, as required by experiments.
Since we are assuming the initial value $\phi_0$ to be of order $H_I$, this means that there must be a sufficient number of efolds between the inital time and the time corresponding to $N_{3000 h^{-1}{\rm Mpc}}$. This can be achieved if the value of $\phi$ at the end of  the {\it phase I} (which is roughly $M/\sqrt{\beta}$) is much larger than $H_I$. Therefore this tells us that $M$ has to be large enough with respect to $H_I$. Precisely we may write $\phi(t) \simeq
M/ \sqrt{\beta} e^{4 \beta(H t - N_I)}$ (valid for small $\beta$) and so the amplitude for the mode corresponding to present horizon, reduces to:
\be
 A_{C} = \frac{H_I^2} {64 \beta \pi^2 M^2} e^{8\, \beta \Delta N_{\rm hor}}\ee
where $\Delta N_{\rm hor} \equiv N_{ 3000\,h^{-1}\,{\rm Mpc}} - N(t_I)$.

In order to connect to constraints form observations let us also introduce the total perturbation in the energy density $\zeta$, which is gauge-invariant and defined as:
\be
\zeta\equiv -\varphi-H\frac{\delta\rho_{\rm tot}}{\dot{\rho}_{\rm tot}}   \label{zetadef} \, ,
\ee
where $\rho_{\rm tot}$ is the total energy density, and $\delta\rho_{\rm tot}$ stands for a perturbation about the average value.
One can compute it in the uniform gauge and find that 
 $\zeta(x)=C(x)$ neglecting the decaying mode in the Eq.~(\ref{CD}) (see
appendix~\ref{Amplitude}). The quantity $\zeta(x)$ is conserved on
superhorizon scales in any cosmological epoch, and therefore it is related to
density perturbations at late times and it is constrained by experiments. Precisely, its amplitude has to be smaller than $4 \times 10^{-10}$, if we want to suppress the spectrum (since its spectral index is not flat).
So this requirement constrains the ratio $H_I/M$ (we give here the result for small $\beta$):
\be
\frac{H_I}{M} \lesssim 16\pi \sqrt{\beta} e^{-4 \beta\Delta
  N_{\rm hor} }  \times 10^{-5}  \, .    \label{suppress}
\ee
This condition guarantees a sufficient number of e-folds
from the initial time $t_0$  until the time at which the scale corresponding to the present horizon is produced.

We can also work out a necessary condition: since the right hand side has a maximum for $\beta=1/(8\Delta N_{\rm hor})$
and $\Delta N_{\rm hor} > 4.3$ \footnote{It is immediate from the second of
  our basic requirements in sect.~\ref{bubble} and from Eqs. (\ref{dueN}).}, we find:
\be
\frac{H_I}{M} < 5 \times 10^{-5} \, , \label{necessary}
\ee
which is also necessary to satisfy the phenomenological bounds on tensor perturbations (see Eq.~(\ref{tensor})).




\subsection{Adding a curvaton} \label{curvaton}



As we said at the beginning of this section, another field
($\sigma$) may be responsible of the generation of the curvature
perturbations. This is necessary (unless we modify the Lagrangian to have a strong transition even when $\beta$ is small during the {\it phase I}, as in~\cite{tirtho}) since the spectral index of perturbations in $\phi$ is not
close enough to $1$. In order to do that we require the field $\sigma$ to be
minimally coupled (or with very small nonminimal coupling) and light. First of all, note that if $H_I$ happens to be small enough (see Eq.~(\ref{suppress})) then the amplitude of the fluctuations
generated by the field $\phi$ alone is negligible. In this case it is the
field $\sigma$ that can give rise to the observed spectrum of perturbations,
through the curvaton mechanism  \cite{lyth}.
So, we consider here a minimally coupled field $\sigma$, whose mass $m_{\sigma}$ is smaller than $H_I$.
In other words we have to add to the initial lagrangian of Eq.~(\ref{lagrangiana}) the following terms:
\be
L_m=-\frac{1}{2}\partial_{\mu}\sigma \partial^{\mu} \sigma - \frac{1}{2}m_{\sigma}^2 \sigma^2  \, ,
\ee
where for simplicity we have chosen a quadratic potential.
Since the field is light it develops quantum fluctuations with an almost scale invariant spectrum during the {\it phase I}, in which $H$ is constant. The spectral index is given by:
\be
n_{\sigma}\simeq 1-\frac{2 m^2_{\sigma}}{3 H^2} \, .
\ee
The reason why it differs from the $\phi$ fluctuations is that fluctuations in
a minimally coupled field are sensitive only to the slow-roll parameters (that
is the evolution of the background), while a non minimally coupled field is
also sensitive to the coupling $\beta$\footnote{Another way to see it is to compute the perturbations in the Einstein frame \cite{tirtho}. In this frame
  the field $\phi$ becomes slowly rolling, with the $\epsilon$ parameter very suppressed in the exponential phase and the $\eta$ parameter proportional to
  $\beta$. So, the spectral index $n-1$ for $\phi$ is proportional to $\eta$ and therefore to $\beta$, while for a curvaton field it is suppressed, since
  it is only proportional to $\epsilon$ and not to $\eta$ \cite{lyth}. For a
  more general analysis of perturbations for a two-field system in the Einstein
  frame, the case of the present lagrangian is also recovered by generalized
  theories of gravity examinated in \cite{staro}.}.

Also, since it has no coupling with $\phi$, its energy-momentum tensor is conserved and so the variable:
\be
\zeta_{\sigma}\equiv -H\frac{\delta \rho_{\sigma}}{\dot{\rho}_{\sigma}}
\ee
is constant on large scales.
In fact, as it was shown in \cite{Wands}, the quantities:
\be
\zeta_{i}\equiv -\varphi - H\frac{\delta\rho_{i}}{\dot{\rho}_{i}}
\ee
(where the subscript ``$i$'' refers to one single component of the energy
content of the Universe) are conserved on large scales if the energy-momentum tensor
of the $i$ component is conserved, irrespectively of
the gravitational field equations. So this applies to generalized metric
theories of gravity and so also to our case.
Now, the total $\zeta$ can be expressed as a combination of the individual $\zeta_i$:
\be
\zeta=-\frac{\zeta_{\rm rest} \dot{\rho}_{\rm
    rest}+\zeta_{\sigma}\dot{\rho}_\sigma  }{ \dot{\rho}_{\rm rest}+ \dot{\rho}_{\sigma}} \, ,
\ee
where the subscript ``rest'' stands for any other component in the Universe.
So, if $\dot\rho_{\sigma}$ evolves differently from  $\dot\rho_{\rm rest}$, then $\zeta$ is no more constant.
In particular, if the numerator is dominated by
$\zeta_{\sigma}\dot{\rho}_\sigma$, then the variable $\zeta$ acquires also a
flat spectrum.

The curvaton scenario uses the fact that when $H$ becomes smaller than $m_{\sigma}$, the field starts oscillating.
In our case we also need that $m_{\sigma}\lesssim \Gamma_{\rm vac}^{1/4}$, so that the oscillations start after the end of inflation $t_E$, so everything proceeds as in the conventional curvaton scenario. Therefore the ``rest'' is the usual radiation component during the radiation era.

In fact, $\sigma$ is frozen to its initial value ($\sigma_I$) before the time
$t_{\rm end}$ (which is the moment at which radiation is created by the nucleation and collisions of bubbles) and then it starts oscillating when $H\simeq m_{\sigma}$. So at this point it starts behaving like matter and eventually it becomes dominant as in the usual curvaton models \cite{lyth}, since it redshifts less fast than radiation. Therefore its $\zeta_{\sigma}\dot{\rho}_\sigma= - 3 H \zeta_{\sigma}\rho_\sigma$ can dominate and so $\zeta\simeq\zeta_{\sigma}$.
In this case the amplitude of the power spectrum is given by:
\be
{\cal P}^{1/2}_{\zeta}= r_C \frac{H_I}{\pi \sigma_I} \, , \label{amplcurv}
\ee
where $r_C$ is the ratio of energy density in the curvaton field to the energy density stored in radiation at the epoch of curvaton decay (it is constrained to be $1 \geq r_C \gtrsim 10^{-2} $), and $\sigma_I$ is the value of the field $\sigma$ during the {\it phase I} of inflation.
The amplitude of Eq.~(\ref{amplcurv}) has to be the observed $10^{-5}$
\footnote{Note that it is reasonable that $\zeta_{\phi}$ (see Eq.~(\ref{powerzeta})) is suppressed with respect to $\zeta_{\sigma}$. In fact, since $\phi_I$ is unstable, it is natural to have $\phi_I\geq\sigma_I$.}.

In order for this mechanism to work, the curvaton has to dominate the universe before the nucleosynthesis epoch, so:
\be
\frac{1}{2} m_{\sigma}^2 \sigma_I^2 \left( \frac{\rm MeV^2}{m_{\sigma} M_{\rm
      Pl}}\right)^{3/2}\gtrsim {\rm MeV}^4 \, .
\ee
Now, the maximal value for $m_{\sigma}$ is $\Gamma_{\rm vac}^{1/4}$ and the  value for $\sigma_I$ is around $10^{5} H_I$, so there is a possibility of making the mechanism work if:
\be
H_I \gtrsim 10^7 {\rm GeV}\,r_0^{-1/20}\, ,
\ee
which generalizes the usual result that
$H_I\gtrsim 10^{7} {\rm GeV}$ in the simplest curvaton scenario \cite{lythbound}.



\subsection{Gravitational waves}

Here we stress that in our scenario gravity waves are created in the usual way on cosmological scales during the phase of exponential expansion.
As long as $\phi$ is small so that we are in the {\it phase I}, the spectrum for the gravity waves is exactly the usual one and exactly scale invariant:
\be
{\cal P}_T= \frac{2 H_I^2}{\pi^2 M^2} \, .
\ee
This is constrained to be smaller than $A^2 \approx 10^{-10}$ by CMB experiments \cite{MAP}, so:
\be
\frac{H_I}{M} \lesssim 10^{-5} \,\, ,  \label{tensor}
\ee
which is consistent with Eq.~(\ref{necessary}).




\section{Late time gravity}  \label{einstein}



Since the parameter $\beta$ has to be big enough (see
Eq.~(\ref{boundnumerico})), it is apparent that the model does not lead
directly to Einstein gravity in the late Universe, but to Brans-Dicke
gravity. In fact an additional scalar may mediate a ``fifth force'' between bodies at the level of the solar system,
spoiling the successful predictions of General Relativity.

We may distinguish two possibilities. If the scalar field is very massive (with respect to the inverse of the solar system lenght) then its influence is negligible in solar system experiments, since it mediates a force with too short range.

If, instead, the scalar mass is small with respect to the inverse solar-system distance, then it must be presently very weakly coupled to matter for the model to be consistent with observational data.
Deviations from General Relativity in scalar-tensor theories are usually parametrized, defining
\be F(\phi)\equiv (M^2+\beta \phi^2) \, ,\ee
by the following Post-Newtonian parameters:
\be
\gamma-1=\frac{ (dF/d{\phi})^2 }   { F+ 2 (dF/d{\phi})^2  }
\qquad , \qquad  \mu-1= \frac{1}{4} \frac{F(dF/d{\phi}) }{2F+3 (dF/d{\phi})^2 }  \frac{d\gamma}{d\phi} \, .
\ee
The limits coming form the experimental bounds on these parameters \cite{reviews} lead to the following constraint:
\be
\frac{(dF/d{\phi})^2}{F}= \frac{4 \beta^2 \phi^2}{M^2+\beta\phi^2} \lesssim  10^{-5}  \, . \label{PNbound}
\ee
This was recognized as a potential problem also in the Extended Inflation
scenario, since in the original Lagrangian there is no mass for the $\phi$ field.
 However the problem is not so hard to tackle, since there is a long
time evolution in the system between the inflationary epoch and the epoch in
which gravity is tested. So, just as for the Extended Inflationary scenario we
can use different possibilities for recovering Einstein gravity at late times.

There are several strategies to overcome this problem:
\begin{itemize}
\item Drive $\phi$ at late times to a value smaller than $M$, so that Eq.~(\ref{PNbound}) is satisfied for any $\beta$.
This is possible for example adding a potential term for $\phi$ that drives it to zero.
In this case $M$ really corresponds to the observed $M_{\rm Pl}$.
In order not to change the previous discussion, the potential $V$ has to satisfy
$
V \ll\ \Lambda
$.
Note that in this case we have Einstein gravity irrespectively of the value of $\beta$, and irrespectively of the fact that the field has a mass today.
\item In a similar way one can imagine a potential $V(\phi)$ (that again has
  to satisfy the condition $V \ll\ \Lambda $ ) that locks the field $\phi$ at
  some generic value (which can be also bigger than $M$) giving it a mass
  bigger than the inverse of the solar system lenght. In this case there may
  well be significant post-inflationary evolution, and the value $M_{\rm Pl}$ today may be significantly different from the value $M$. The fifth force experiment constraints are avoided since the field is massive today.
\item
Another strategy is to modify the model (as in hyperextended inflation
\cite{hyperextended}, or in \cite{Barrow-Maeda,gbellidoquiros}), in such a way that $\beta$ is not a constant, but it can vary with time.
In this case it is sufficient to have a dynamics such that $\beta$ can end up having a very small value at late times, so that Eq.~(\ref{PNbound}) is satisfied.
\item Finally another interesting strategy is to couple $\phi$ differently to
  the matter sector (as in \cite{holman}) and to the vacuum energy. This is a
  clear procedure in the so-called Einstein frame (in which gravity is
  described by the usual $R M_{\rm Pl}^2$ term only). In our frame \cite{holman} the modification consists in substituting the $\Lambda$ term with:
\be
S_{\Lambda}= - \frac{1}{2} \int d^4x \sqrt{-g} \, \Lambda \left(\beta
  \frac{\phi^2}{M^2}\right)^{2(1-\eta)}\, .
\ee
The addition of the new parameter $\eta$ (introduced by the generalized Brans-Dicke model of~\cite{Damour}) makes possible to satisfy all the constraints derived in section \ref{bubble} without contradicting current experiments on gravity.
\end{itemize}

Models with a potential are simple possibilities, but on the other hand one may argue that they introduce small parameters. It is true in fact that, in the sense of \cite{Freese}, the potentials have to be very flat: if one computes the ratio $\Delta V/(\Delta\phi)^4$ (where $\Delta V$ is the change in potential energy of the field during the whole duration of inflation, and $\Delta\phi$ is the variation of the field) one finds that very small values are required.
However we think that this is not fine-tuning, since the only physical requirement is that $V\ll \Lambda$ (where a small hierarchy is sufficient): then the fact that $\Delta\phi^4$ can become much bigger than $V$ is a result of the dynamics and it is not put by hand.



\section{Other consequences and possible effects } \label{other}



In this section we mention some of the observational consequences of the model that have to be explored and that are not covered in this paper.

One is the production of bubbles on the small scales, which could be a striking signature of the model.
If bubbles are produced at small scales, these could be detected as additional power in the CMB spectrum at high multipole ($l\gtrsim 1000$), and as presence of many voids on the small scale of the galaxy distributions.
Already a few groups have investigated in this direction \cite{Sakai}-\cite{Amendola}, and we will explore in future work the possible imprints of our specific model in the observations on small scales.

A second consequence is that our model can automatically incorporate a change of the spectral index at some small scale, which might be necessary to fit the data as suggested by the WMAP \cite{MAP}. In fact the spectral index of any fluctuating field becomes smaller at the scale corresponding to the transition form the exponential inflation to the power-law inflation.

Another effect to be computed is the non-gaussianity in the cosmological perturbations produced in this specific model: for example in case these are produced by a curvaton this is likely to be relevant. Finally also the production of gravity waves from Bubble collisions might be potentially observable.

\vskip 0.5cm

Also, we mention here some possible variations on the model and some points that are still missing in the analysis of this paper.

First of all, in the presence of a time dependent background metric, the decay
rate $\Gamma_{\rm vac}$ could acquire a time dependence
\cite{falsevacuum}. This could lead to a different constraint on the
parameters of the theory. In particular the constraint on $\beta$, and thus on
the spectral index of the $\phi$ fluctuations can be different (for example an
order of magnitude difference in $\Gamma_{\rm vac}$ at the beginning of
inflation and $\Gamma_{\rm vac}$ during the asymptotic stage leads to a $0.02$ difference in the spectral index). This effect will be subject to further study.

Then, there are ways to make the model viable even with the field $\phi$ alone. One might in fact invoke a different coupling of the field $\phi$ with the Ricci scalar.
Instead of the $\beta R \phi^2$ coupling one may consider for example:
\be
 \kappa \, R \, \ln \left[\frac{R}{M^2}\right] \phi^2 \, .
\ee
This was proposed in~\cite{Ford}, where the author suggested that such a term in the Lagrangian can arise through quantum corrections. In this case the cosmological dynamics is described by an effective time dependent quantity:
\be
\beta_{\rm eff}= \kappa \ln\left[\frac{R}{M^2}\right] \,  ,
\ee
where $\kappa$ is a number. This is justified as long as the variation in
$\beta_{\rm eff}$ is sufficiently slow that its time derivatives may be neglected.
As $R$ decreases, $\beta_{\rm eff}$ incresases thus making the process more efficient. In this way the bound on the spectral index of the perturbations of the field $\phi$ gets relaxed.

The same thing applies for any quantum correction to the parameter $\beta$, that could change the bounds in a relevant way.

Generally speaking our idea of having exponential inflation slowed down by an unstable scalar field could be easily implemented in other variants of the model, which make the transition stronger so that the curvaton becomes unnecessary \cite{tirtho}. Basically wath is needed is to have a coupling with $R$ that grows with $\phi$, faster than the $\phi^2 R$ coupling analyzed in the present paper.



\section{Conclusions} \label{conclusions}



We have shown in this paper that a false vacuum can successfully decay to a true vacuum, producing inflation, in the presence of a non minimally coupled scalar $\phi$, since Exponential Inflation is slowed down to Power-Law Inflation.

Then we have analyzed the constraint coming from the fact that we do not want to produce large bubbles that would spoil the CMB.
This implies the following constraints: $r_0=\Gamma_{\rm vac}/H_I^4 \lesssim
10^{-7}$ (where $\Gamma_{\rm vac}$ is the tunneling rate per unit volume of the false vacuum to the true vacuum, and $H_I$ is the Hubble constant during de Sitter Inflation) and the nonminimal coupling has to be $\beta \gtrsim 1/57$.

If $\phi$ is responsible also for primordial density fluctuations then
the spectral index is $n_S\simeq1-8\beta \lesssim 0.86$, which is in disagreement with observations.

However, if there is also a minimally coupled scalar, it can produce a flat spectrum through the curvaton mechanism.

Moreover, generally speaking the same idea can easily be adapted modifying slightly the model in order to have a stronger transition to Power-Law inflation, and so without the need for a curvaton field. One example \cite{tirtho} is given by generalizing the coupling $\phi^2 R$ to a generic function $f(\phi) R$, where $f(\phi)$ grows faster than $\phi^2$ for large $\phi$.
Another example \cite{Ford} is to couple a slightly different function of $R$ to $\phi^2$ (see section~\ref{other}).

Finally the model can lead to other observable consequences (as discussed in section~\ref{other}), as a change in the spectral index (``running'' spectral index) at some small scale , the production of bubbles on the small scales (detectable as voids in the large scale structure and at large $l$ in the CMB) and possibly the presence of some non-gaussianity. All these effects are subject to future work.



\section*{Acknowledgments}

We are grateful to T.~Biswas, R.~Brandenberger, R.~Catena, F.~Finelli and B.~Katlai for useful discussions and comments.




\appendix

\section{Bubbles on the CMB} \label{BubblesApp}

Here we follow \cite{LiddleWands} to estimate the constraint on
$r_0=\Gamma_{\rm vac}/H_I^4$, using the WMAP data.

As we have mentioned, the presence of a bubble would be directely detected as a void region (so a region with $\delta\rho/\rho=-1$) in the CMB, if the bubble size (whose comoving value will be called $r$) is large enough with respect to the resolution (corresponding to a comoving size $r_b$) of the experiment.
The fluctuation in the temperature (through a fluctuation in the Newtonian gravitational potential $\Phi$) that it would produce is:
\be
\frac{\delta T}{T}\simeq-\frac{\Phi}{3}\simeq\frac{1}{3}
\left(\frac{\delta\rho}{\rho}\right) \left(\frac{r}{L_{\rm lss}} \right)^2  \, ,
\ee
where $\delta\rho/\rho=-1$ for voids, and where $L_{\rm lss}$ is the comoving horizon size at last scattering (and the Doppler contribution has been neglected as in \cite{LiddleWands}).
If the bubble size is smaller than the resolution, it can still be detected, but the $\delta T/T$ is reduced by a factor
$(r/r_{b})^2$.
Also, if a bubble is smaller than the last scattering surface width (whose comoving value we call $r_w$), it is reduced by a factor $(r/r_w)^2$.
So, the minimal bubble size that can be detected is:
\be
r_{\nu}\approx \left(3 \times 10^{-5} L^2_{\rm lss} r^2_b r^2_w
\right)^{1/6}\approx  6 {\rm Mpc} \, ,
\ee
where we have used the following values: $ L_{\rm lss}\simeq 190 h^{-1} {\rm Mpc}$,
$r_b\approx 7 h^{-1}{\rm Mpc}$ and $r_w\simeq 6.7 h^{-1}{\rm  Mpc}$.
Any bubble which produces a fluctuation bigger than what is observed (some $10^{-5}$) is in contradiction with observations.

Note that the pixel size of the WMAP experiment is 30 times smaller than the
COBE's~\cite{MAP}, but we have used only the data up to $l\approx 600$ (which
corresponds to a comoving scale of about $7 h^{-1} {\rm Mpc}$), for which the signal-to-noise ratio is less than 1. Such a small resolution is what makes the bound stronger today, with respect to the numbers used by~\cite{LiddleWands}.

The requirement in order to be safe from seeing big bubbles is that we demand that the number of voids inside the horizon be less than that which gives a 95\% confidence level that at least one is in the last scattering surface. This is true if:
\be
\frac{1}{3} \int_{r_{\nu}}^{\infty} \frac{4 \pi L_H^2 r}{(4\pi/3)L_H^3} \left( \frac{dN_B}{dr} \right) dr <1 \label{constraint}  \, ,
\ee
where $\frac{dN_B}{dr}$ is the spectrum of bubbles ($N_B$ is the number of bubbles) which is generated by a specific model and where $L_H=3000 h^{-1}{\rm Mpc}$ is the present horizon distance.

This spectrum, in our model, is calculated as follows.

A bubble nucleated at time $t_1$ with zero initial size grows at the speed of light within the expanding universe and it has size $x$ at time $t$:
\be
x(t,t_1)=a(t) \int_{t_1}^{t} \frac{dt}{a(t)} \, .
\ee
At time $t_1$ the bubble nucleation probability per unit comoving volume is:
\be
\frac{dN_B}{dt_1}=\Gamma_{\rm vac} \, a(t_1)^3 \, .
\ee
We can change variables to $x$ and evaluate the spectrum at the end of
inflation ($t_{\rm end}$):
\be
\frac{dN_B}{dx}=\Gamma_{\rm vac} \, a(t_{\rm end})^3 \frac{a(t_1)^3}{a(t_{\rm end})^3}\,\frac{dt_1}{dx}  \, ,
\ee
and then, compute it in a sphere whose radius $l_H$ corresponds to the physical size of our present horizon at that time ($\frac{4 \pi}{3} l_H^3$) :
\be
\frac{dN_B}{dx}=r_0 H_I^4 \,\frac{4 \pi}{3}\, l_H^3 \frac{a(t_1)^3}{a(t_{\rm end})^3}\frac{dt_1}{dx}  \, .
\ee
The evolution of the scale factor $a(t)$ (and so the ratio $\frac{dt_1}{dx}$) could be evaluated numerically.
However, we can use a rough approximation that already gives us a correct estimate: we assume that $H(t)$ is constant until it reaches the value $\alpha/t$ and then the scale factor goes as a power law $t^{\alpha}$.
Using this we obtain the following spectrum:
\be
\frac{dN_B}{dx}=-\frac{4 \pi r_0}{3}\frac{H_I^4 H_{\rm end}^4 l_H^4 (\alpha-1)^4}{\left( (H_I/H_{\rm end})^{\alpha}-H_I(\alpha-1) H_{\rm end} x \right))^4}  \, ,
\ee
where we considered values of $x$ that are much bigger than the horizon at the end of inflation ($H_{\rm end}$).

Then we impose a correction factor ($g$) due to the fact that the voids expand faster than the background, and in principle there is also a correction due to the fact that voids get filled by relativistic matter (so their size is reduced by a lenght $\Delta$), so the true comoving lenght of a bubble is:
\be
r=g \left(\frac{L_H}{l_H} x-\Delta \right)  \, .
\ee
However, in a Cold Dark Matter scenario matter becomes nonrelativistic very early, so the void filling lenght $\Delta$ is negligible.
On the other hand, in a CDM dominated universe the void growth is non-negligible \cite{LiddleWands} and it is:
\be
g\simeq 1.85 h^{2/5} \, .
\ee

At this point we can extract the quantity in which we are interested:
\be
\frac{dN_B}{dr}=-\frac{4 \pi r_0}{3}\frac{1}{g L_H}  \left( \frac{L_H (\alpha-1)}{\frac{r_0^{(\alpha-1)/4} L_H}{l_H H_{\rm end}} -(\alpha-1) r/g} \right)^4 \, .
\ee
Note now that the quantity $l_H H_E$ is very big:
\be
l_H H_{\rm end} \approx 4\times 10^{25} \left( \frac{T_{\rm rh}}{10^{15} h \,{\rm GeV}} \right) \, ,
\ee
so we may safely neglect the first term in the denominator and we get the very simple result:
\be
\frac{dN_B}{dr}=-\frac{4 \pi r_0}{3}\frac{g^3 L_H^3}{r^4} \, .
\ee
Finally, by integrating this in Eq.~(\ref{constraint}) we obtain the following:
\be
r_0 \lesssim 10^{-7} \, .
\ee


\section{Amplitude of $\phi_0$ and $\zeta$} \label{Amplitude}


It is interesting  to understand on physical grounds what should be the value for
$\phi_0$. If we put for example exactly $\phi_0=0$ as initial value (so that  classically $\phi=0$ would be an (unstable) solution of the equations of motions at all times), we would find that the amplitude
of the fluctuations of $C$ diverges for $\phi_0=0$.
However the fluctuation $\delta\phi$ is not divergent, and so the value that $\delta\phi$ assumes in one Hubble patch should be included into the initial classical value $\delta\phi_0$.
In other words the minimal value of $\phi_0$ should be taken as given by the typical value of the quantum fluctuation of the field.
This value can be taken from the expression:
\be
|\delta\phi(x)|^2=\int_{k_{\rm min}}^{k_{\rm max}} \frac{dk}{k} {\cal
  P}_{\delta\phi_{\varphi}}=\left(\frac{\dot{\phi}}{H}\right)^2\int_{k_{\rm
    min}}^{k_{\rm max}} \frac{dk}{k} {\cal P}_{C}   \label{phi_0}  \, .
\ee
As $k_{\rm min}$ we choose $k_{\rm min}=a_0 H$, the wavelenght produced at the beginning of the inflationary phase.
As $k_{\rm max}$ we choose instead $k_{\rm min}=a(t_l) H$, the last produced wavelenght at the time defined by $a(t_l)$.
As a result, we find:
\be
|\delta\phi(x)|^2=\frac{2^{\epsilon}}{\pi^3} \left| \Gamma(3/2+\epsilon/2) \right|^2 \frac{H^2}{\epsilon} \left[ \left(\frac{a_l}{a_0}\right)^{\epsilon}-1\right]  \, .
\ee
Substituting $\frac{a_l}{a_0}=e^{N_{\rm tot}-N_l}$ we get:
\be
|\delta\phi(x)|=\frac{2^{\frac{\epsilon}{2}}}{\pi^{3/2}}
 \Gamma(3/2 + \epsilon/2)\frac{H}
{\sqrt{\epsilon}} \sqrt{e^{(N_{\rm tot}-N_l)\epsilon}-1 } \, .\ee
This means that the field stays constant for about $2/\epsilon$ e-folds and then it starts to grow.
So the physical initial value that the field takes (assuming that classically the field is initially set to zero) is the value of the quantum fluctuations after about $2/\epsilon$ e-folds from the beginning.
This means that the minimal initial condition is given by:
\be
\phi_0\approx 2 H_I \frac{2^{\frac{\epsilon}{2}}\Gamma(3/2+\epsilon/2)}{\sqrt{\epsilon}}
\qquad
\left(\phi_0 \approx \sqrt{\frac{\pi }{8 \beta}} H_I   \hspace{0.3cm} \, \,
{\rm for \, small}\,\, \beta \right)  \, . \label{fi0}
\ee

Going back to the computation of the amplitude of the perturbations, the last step consist in relating the spectrum of $C$ to a conserved quantity during the post inflationary evolution.
One possibility is to use the variable defined in Eq.~(58) of \cite{Hwang2}, which is conserved superhorizon and it is equal to $C$.

Another possibility is to use the variable $\zeta$ that we have defined in
Eq.~(\ref{zetadef}), where we may insert $\delta\rho_{\rm tot}$ and
$\dot{\rho}_{\rm tot}$ following \cite{Hwang}. In particular $\delta\rho_{\rm
  tot}$ corresponds to Eq. (33) of \cite{Hwang} where one has to insert the
solutions of the system of Eqs. (54-56) of that work.
In fact the energy density and pressure defined in Eqs. (30, 31) coincide with
$\rho_{\rm tot}$ and $p_{\rm tot}$ and they are conserved, if one sets the potential $V$ defined in those equations equal to a constant $\Lambda$:
\begin{eqnarray}
\rho_{\rm tot}&=&\frac{M^2}{F}\left( \frac{\dot{\phi}^2}{2}+\Lambda -3 H \dot{F}\right)=\rho_{\phi}+\Lambda  \, ,\\
p_{\rm tot}&=&\frac{M^2}{F} \left(  \frac{\dot{\phi}^2}{2} -\Lambda +\ddot{F}+2 H\dot{F}\right)= p_{\phi}-\Lambda \, ,
\end{eqnarray}
where $F\equiv M^2+\beta \phi^2$.
One can verify explictly that they obey the continuity equation
$\dot{\rho}_{\rm tot}=-3 H (\rho_{\rm tot}+p_{\rm tot})$.

So, one can express $\zeta$ as a function of $\delta\phi$ and therefore as a
function of $C$, finding (after a lenghty computation) that $ \zeta=C $.



\appendix











\end{document}